\input amssym.def
\input amssym.tex

\def\Z{\Bbb Z}\def\R{\Bbb R}
\def\hfb{\hfill\break}

\input harvmac
\magnification 1200 
\hsize=16.5truecm 
\vsize=23.5truecm 
\overfullrule=0pt
\parskip=3pt
 at 14truept
\overfullrule 0pt


\def\hexnumber@#1{\ifcase#1 0\or1\or2\or3\or4\or5\or6\or7\or8\or9\or
        A\or B\or C\or D\or E\or F\fi }

%
\def\boxit#1{\leavevmode\kern5pt\hbox{
	\vrule width.2pt\vtop{\vbox{\hrule height.2pt\kern5pt
        \hbox{\kern5pt{#1}\kern5pt}}
      \kern5pt\hrule height.2pt}\vrule width.2pt}\kern5pt}



\def\\{\hfil\break}

\def\b#1{\big(#1\big)}

\def\+{\oplus}
\def\x{\otimes}
\def\wt{{\rm wt}}

\def\b#1{\kern-0.25pt\vbox{\hrule height 0.2pt\hbox{\vrule
width 0.2pt \kern2pt\vbox{\kern2pt \hbox{#1}\kern2pt}\kern2pt\vrule
width 0.2pt}\hrule height 0.2pt}}

\def\STrow#1{\hbox{#1}\kern-1.35pt}

\font\huge=cmr10 scaled \magstep2
\font\small=cmr8


{\nopagenumbers
\ \ 
\vskip 2cm
\centerline{{\huge\bf On the level-dependence of Wess-Zumino-Witten
three-point functions}}
\bigskip
\bigskip\bigskip\centerline{J\o rgen Rasmussen {\footnote{$^\ddagger$}
{\small supported
in part by a PIMS Postdoctoral Fellowship and by NSERC. E-mail:
rasmussj@cs.uleth.ca}} \quad\quad\quad Mark A. Walton
{\footnote{$^\dagger$}{\small supported in part by NSERC. E-mail:
walton@uleth.ca}}} \bigskip

\centerline{{\it Physics Department,
University of Lethbridge}} \centerline{{\it Lethbridge, Alberta,
Canada\ \ T1K 3M4}}

\vskip 1.5cm \leftskip=1cm \rightskip=1cm
\centerline{{\bf Abstract}}

\noindent 
Three-point functions of Wess-Zumino-Witten models are investigated. In
particular, we study the level-dependence of three-point functions in the
models based on algebras $su(3)$ and $su(4)$. We find a correspondence 
with Berenstein-Zelevinsky triangles. 
Using previous work connecting those triangles to the fusion multiplicities, 
and the Gepner-Witten depth rule,  
we explain how to construct the full three-point functions.  
We show how their level-dependence 
is similar to 
that of the related fusion multiplicity. For example, the concept of threshold
level plays a prominent role, as it does for fusion. 

\leftskip=0cm \rightskip=0cm

\vskip1cm


\vfill
\eject}

\pageno=1

\lref\kmswii{A.N. Kirillov, P. Mathieu, D.
S\'en\'echal, M.A. Walton, in: Group-Theoretical Methods in Physics,
Proceedings of
the XIXth  International Colloquium, Salamanca, Spain, 1992, Vol. 1
(CIEMAT, Madrid, 1993)}

\lref\jrfree{J. Rasmussen, 
Applications of free fields in 2D current
algebra, Ph.D. thesis (Niels Bohr Institute, 1996), hep-th/9610167;
J.L. Petersen, J. Rasmussen, M. Yu, Nucl. Phys. {\bf B502} (1997) 649}

\lref\cmw{C.J. Cummins,
P. Mathieu, M.A. Walton, Phys. Lett. {\bf 254B} (1991)
386}\lref\kmsw{A.N. Kirillov, P. Mathieu, D. S\'en\'echal, M.A. Walton,
Nucl. Phys. {\bf B391} (1993) 651}\lref\bmw{L. B\'egin, P. Mathieu, 
M.A. Walton, Mod. Phys. Lett. {\bf A7} (1992) 3255}\lref\bkmw{L. B\'egin, 
A.N. Kirillov, P. Mathieu, M.A. Walton, Lett. Math. 
Phys. {\bf 28} (1993) 257} 

\newsec{Introduction}

A general study of Wess-Zumino-Witten
(WZW) three-point functions was initiated in 
 \ref\jrthree{J. Rasmussen, Int. J. Mod.
Phys. {\bf A14} (1999) 1225}. At sufficiently high
level $k$, three-point functions for so-called generating-function
primary fields were written as
\eqn\iiipt{\langle\, \phi_\lambda(z_1,x_1)\phi_\mu(z_2,x_2)\phi_\nu(z_3,x_3)
\,\rangle\ =\ \Delta(z_1,z_2,z_3)\, 
F^{(k)}_{\lambda,\mu,\nu}(x_1,x_2,x_3)\ .} 
A (generating-function) primary field $\phi_\lambda(z,x)$ is labelled by a 
dominant weight $\lambda$ in 
\eqn\Pk{P_\ge^k\ =\ \big\{\sigma =\sum_{j=1}^r\,\sigma_j\Lambda^j\,|\, 
\sigma_j\in \Z_\ge,\,\forall j=1,\ldots,r;\, 
\,\sum_{j=1}^r\sigma_ja^{\vee j}\le k\,\big\}\ ,} 
so that its dependence on the level $k$ is implicit. 
Here $\Lambda^j$ is the $j$-th fundamental weight of a simple Lie algebra
$X_r$, of rank $r$, and $a^{\vee j}$ is the corresponding co-mark. That is, 
the highest root of $X_r$ has co-root $\theta^\vee\ =\ \sum_{j=1}^r 
a^{\vee j} \alpha_j^\vee$, where $\alpha_j^\vee$ is co-root to the $j$-th 
simple root $\alpha_j$. Henceforth, we will normalise the long roots 
$\alpha_{\rm long}^2=2$, so that $\theta^\vee=\theta$.  

In 
$\phi_\lambda(z,x)$, $z$ is the holomorphic world-sheet coordinate, and 
$x$ represents the flag variables - there is an independent 
variable $x^\alpha$ for every positive root $\alpha\in R_>$ of the simple 
Lie algebra $X_r$.

In \iiipt, \eqn\zdelta{\Delta(z_1,z_2,z_3)\ :=\ 
(z_1-z_2)^{-\Delta_1-\Delta_2+\Delta_3}\, 
(z_2-z_3)^{\Delta_1-\Delta_2-\Delta_3}\, 
(z_3-z_1)^{-\Delta_1+\Delta_2-\Delta_3}\ ,}
where $\Delta_j,\ j=1,2,3$, denote the conformal weights of the three 
primary fields. 
That is, the primary field $\phi_\lambda(z_1,x_1)$ obeys the following 
operator product expansions (OPEs) with the energy-momentum 
tensor $T(z)$
\eqn\gfprimi{T(z)\,\phi_\lambda(z_1,x_1)\ \sim\ 
{{\Delta_1}\over{(z-z_1)^2}}\phi_\lambda(z_1,x_1)
\ +\ {1\over{z-z_1}}{{\partial\phi_\lambda(z_1,x_1)}
\over{\partial z_1}}\ ,}
and the currents $J_a(z)$
\eqn\gfprimii{J_a(z)\,\phi_\lambda(z_1,x_1)\ \sim\ 
{{-1}\over{z-z_1}}{\omega(J_a)(x_1,\partial,\lambda)}\phi_\lambda(z_1,x_1)\ .} 
Similar OPEs are obeyed by  $\phi_\mu(z_2,x_2)$ and $\phi_\nu(z_3,x_3)$. 
$J_a(x_1,\partial,\lambda)$ denotes a differential operator (in 
$x_1$) that realises the corresponding finite-dimensional Lie algebra 
generator $J_a$ - see \refs{\jrthree{,}\jrfree} 
for more details. In 
\gfprimii, $\omega(J_a)$ is the generator of $X_r$ obtained from $J_a$  by 
the Chevalley involution $\omega$. 
Since the conformal weights depend on the level, the function 
$\Delta(z_1,z_2,z_3)$ has an implicit and well-known dependence on the level. 

Here we investigate the extension of \iiipt\ from high level 
to all levels $k\in\Z_\ge$. We will treat different levels 
together, investigating the so-called level-dependence of the 
three-point functions. The implicit level-dependence of the factor 
$\Delta(z_1,z_2,z_3)$ is well-known, so the object of study will be 
$F^{(k)}_{\lambda,\mu,\nu}(x_1,x_2,x_3)$. 

Section 2 is a general discussion of three-point functions. It also contains 
a review of threshold levels in fusion and indicates how they  
enter consideration of three-point functions. Sections 3 and 4 treat the 
cases $su(3)$ and $su(4)$, respectively. (The case of $su(2)$ is 
covered in Appendix A.) Section 5 is a concluding discussion.

\newsec{Three-point functions}

It is important here that the Ward identities can be 
written in terms of the differential operators $J_a(x,\partial,\lambda)$. 
For example, we have 
\eqn\wardf{0\ =\ \bigg(\,\sum_{j=1}^3\,J_a(x_j,\partial,\Lambda_{(j)})
\,\bigg)\,\,
F^{(k)}_{\Lambda_{(1)},\Lambda_{(2)},\Lambda_{(3)}}(x_1,x_2,x_3)\ .}
Here we used $\Lambda_{(1)}:=\lambda$, $\Lambda_{(2)}:=\mu$, 
$\Lambda_{(3)}:=\nu$, for convenience of notation. 

Ward identities encoding the level-dependence, however, are not 
imposed by \wardf. Those were used by Gepner and Witten in the derivation 
of their depth rule \ref\gepwit{D. Gepner,  E. Witten,
Nucl. Phys. {\bf B278} (1986) 493}. Let $E_\theta$ denote  
the generator of $X_r$ that raises the weight by the 
highest root $\theta$ of $X_r$.  
In the form 
useful to us, a symmetrical version of the Gepner-Witten constraint is 
\eqn\gwsym{\eqalign{0\ =\ \prod_{j=1}^3\, 
\bigg(\, E_\theta(x_j, \partial, &\,\Lambda_{(j)})\, \bigg)^{\ell_j}\, 
F^{(k)}_{\Lambda_{(1)},\Lambda_{(2)},\Lambda_{(3)}}(x_1,x_2,x_3)\ ,\cr  
&\forall\ (\ell_1,\ell_2,\ell_3)\in \Z_\ge^3,\ 
{\rm such\ that}\ \ell_1+\ell_2+\ell_3 > k\ .}} 

The differential operator expression for $E_\theta$ is particularly simple: 
\eqn\dixth{ E_\theta(x, \partial, \lambda)\ =\ 
{\partial\over{\partial x^\theta}}\ .} 
As a consequence, the differential operator realisation 
is ideally suited to the implementation of the Gepner-Witten depth 
rule:  \gwsym\ takes a simple, useful form,
\eqn\gwsdi{\eqalign{0\ =\ \prod_{j=1}^3\, 
\bigg(\,{\partial\over{\partial x_j^\theta}}\, \bigg)^{\ell_j}\, 
&F^{(k)}_{\Lambda_{(1)},\Lambda_{(2)},\Lambda_{(3)}}(x_1,x_2,x_3)\ ,\cr 
&\forall(\ell_1,\ell_2,\ell_3)\in \Z_\ge^3,\ {\rm such\ that}\ 
\ell_1+\ell_2+\ell_3 > k\ .}} 
This says that the level $k$ must be greater than or equal to the 
highest power 
of $x^\theta$ in $F^{(k)}_{\lambda,\mu,\nu}$, 
using $x_j^\theta=x^\theta$ for all $j=1,2,3$. 

Now $J_a |0\rangle\rangle = 0$, for all generators $J_a$ of the simple 
Lie algebra, where $|0\rangle\rangle$ represents the scalar state 
for the diagonal subalgebra of $X_r\oplus X_r\oplus X_r$. 
From this, we will see shortly that 
\eqn\FG{F^{(k)}_{\lambda,\mu,\nu}(x_1,x_2,x_3)\ =\ 
\left(\, \bigotimes_{j=1}^3\, \langle \Lambda_{(j)} |\, G_+(x_j)\,P\,\right)\ 
|0\rangle\rangle }
satisfies the Ward identities \wardf. Here we follow the notation of 
\jrthree, so that 
\eqn\Gplus{G_+(x)\ :=\ 
\exp\left(\, \sum_{\alpha\in R_>}x^\alpha E_\alpha\,\right)\ ,}
with $E_\alpha$ being the raising operator of $X_r$ associated to 
$\alpha\in R_>$. $P$ is the projector from the algebra 
$X_r\oplus X_r\oplus X_r$ to its diagonal subalgebra. 
Because 
\eqn\wFi{0\ =\ \bigotimes_{j=1}^3\, \langle \Lambda_{(j)} |\,  
G_+(x_j)\, P\  
\left( J_a |0\rangle\rangle \right)\ , } 
we get 
\eqn\wFii{\eqalign{0\ =&\ \bigg(\, \big(\,\bigotimes_{j=1}^3\, 
 \langle \Lambda_{(j)} |\, 
G_+(x_j)\,\big)\, \Delta(J_a)\, \bigg)
\ |0\rangle\rangle\ \cr
=&\ \sum_{\ell=1}^3\,  J_a(x_\ell,\partial,\Lambda_{(\ell)})\, 
\bigotimes_{j=1}^3\,  \langle \Lambda_{(j)} |\, 
G_+(x_j)\,
\, |0\rangle\rangle\ , 
}}
using $P J_a = \Delta(J_a) P$ with 
$\Delta(J_a) = (J_a\otimes I\otimes I) \+ (I\x J_a \x I) \+
(I\x I\x J_a)$.  
So \FG\  obeys \wardf, but 
the condition \gwsym\ must 
still be imposed. 

Let $L(\lambda)$ denote the representation of 
$X_r$ of highest weight $\lambda$. 
We can decompose the polynomial \FG\ as
\eqn\FWa{F^{(k)}_{\lambda,\mu,\nu}(x_1,x_2,x_3)\,=\,
\left(\,\sum_a W_{\lambda,\mu,\nu}^{[a]}(x_1,x_2,x_3)\,
\langle\langle 0|_a\,\right) 
\,|0\rangle\rangle\,=\, \sum_a W_{\lambda,\mu,\nu}^{[a]}(x_1,x_2,x_3)\ ,
}
where $\langle\langle 0|_a$ denotes one of the (normalised) 
singlet states in the triple 
tensor product $L(\lambda)\otimes L(\mu)\otimes L(\nu)$. The $k$-dependence 
will enter in the summation range (see (2.15) below). 

Incidentally, we note that 
$W^{[a]}_{\lambda,\mu,\nu}(x_1,x_2,x_3)$ is a generating function for 
the Clebsch-Gordan coefficients of a coupling 
$L(\lambda)\otimes L(\mu)\otimes L(\nu) \supset L(0)$. We can write 
\eqn\vacC{|0\rangle\rangle_a\ =\ \sum_{|u\rangle\in L(\lambda)}\, 
\sum_{|v\rangle\in L(\mu)}\, \sum_{|w\rangle\in L(\nu)}\,
C_{u,v,w}^{[a]}\, |u\rangle\otimes 
|v\rangle\otimes |w\rangle\ ,} 
where $C_{u,v,w}^{[a]}$ denotes the Clebsch-Gordan coefficient 
appropriate to the coupling indicated by $a$. Define  
\eqn\Kpoly{K_u^\lambda(x)\ :=\ \langle\lambda| G_+(x) |u\rangle\ \ ,}
for $|u\rangle\in L(\lambda)$. 
Then \FG\ and \FWa\ 
yield
\eqn\CGgf{W^{[a]}_{\lambda,\mu,\nu}(x_1,x_2,x_3)\ =\ 
\sum_{|u\rangle\in L(\lambda)}\, 
\sum_{|v\rangle\in L(\mu)}\, \sum_{|w\rangle\in L(\nu)}\ 
C_{u,v,w}^{[a]}\ K^\lambda_u(x_1)\, K^\mu_v(x_2)\, 
K_w^\nu(x_3)\ \ .}

A key step in \jrthree\ was to write the polynomials 
$F^{(k)}_{\lambda,\mu,\nu}(x_1,x_2,x_3)$ as sums of products of elementary 
polynomials:
\eqn\factor{F_{\lambda,\mu,\nu}^{(k)}(x)\ 
=\ \sum_a\,  \prod_{E\in{\cal E}}\ 
\big[\,R^E(x)\,\big]^{p_a(E)}\ .} 
Here $\cal E$ denotes the set of elementary (three-point) couplings of the 
algebra $X_r$, $R^E(x)$ the corresponding polynomial, 
and $(x)$ is short for 
$(x_1,x_2,x_3)$. One necessary constraint on the products 
appearing in the decomposition \factor\ is determined by the weight 
$\{\lambda,\mu,\nu\}$. If we define $\wt(E)$ to be the 
weight of an elementary coupling $E$, then  
\eqn\wtsum{\sum_{E\in{\cal E}}\, p(E) \wt(E)\ =\ \{\lambda,\mu,\nu\} }
must hold. The sets ${\cal E}$ must be found for each algebra 
(see \ref\bcmi{L. B\'egin, C. Cummins, P. Mathieu, J. Math. Phys. {\bf 41} 
(2000) 7611}, for example). There are also 
certain algebraic 
relations among  the elementary couplings (and so among the elementary 
polynomials) that must be taken into account. These relations are sometimes 
called syzygies. They can be implemented by excluding certain products as 
redundant from the sum 
in \factor.  

Once the syzygies are implemented, 
each summand in \factor\ counts a coupling between the primary fields 
$\phi_\lambda(z_1,x_1)$, $\phi_\mu(z_2,x_2)$ and $\phi_\nu(z_3,x_3)$. That 
is, each summand contributes 1 to the fusion multiplicity 
$T_{\lambda,\mu,\nu}^{(k)}$ that corresponds to the three-point function 
\iiipt. 
We therefore rewrite \FWa\ as
\eqn\factorW{F_{\lambda,\mu,\nu}^{(k)}(x)\ 
=\ \sum_{a=1}^{T_{\lambda,\mu,\nu}^{(k)}}\ 
W^{[a]}_{\lambda,\mu,\nu}(x)\ \ ,}
with 
\eqn\WR{W^{[a]}_{\lambda,\mu,\nu}(x)\ =
\ \prod_{E\in{\cal E}}\ 
\big[\,R^E(x)\,\big]^{p_a(E)}\ .}
Since each factor $R^E(x)$ is itself a valid polynomial, 
the $X_r$-symmetry is respected automatically. 
Incidentally, here one also sees the usefulness of the $x$-dependence of the 
generating function primary fields: it makes the different summands 
$W^{[a]}_{\lambda,\mu,\nu}(x)$ independent.  
    
The $k$-dependence of a fixed $W^{[a]}_{\lambda,\mu,\nu}(x)$ 
is straightforward, since by \gwsdi\ we must have
\eqn\gwsW{\eqalign{0\ =\ \prod_{j=1}^3\, 
\bigg(\,{\partial\over{\partial x_j^\theta}}\, \bigg)^{\ell_j}\, 
&W^{[a]}_{\Lambda_{(1)},\Lambda_{(2)},\Lambda_{(3)}}(x_1,x_2,x_3)\ ,\cr 
&\forall\ (\ell_1,\ell_2,\ell_3)\in \Z_\ge^3\ ,\ 
{\rm such\ that}\  \ell_1+\ell_2+\ell_3 > k\ .}}

\subsec{Threshold levels}

Let  $t_a$ denote the maximum power of $x^\theta$ in 
$W^{[a]}_{\lambda,\mu,\nu}(x)$ with non-zero coefficient:
\eqn\taW{\eqalign{t_a\ =\ \min\,\bigg\{\, t\in \Z_\ge\ |\  \prod_{j=1}^3\, 
\bigg(\,{\partial\over{\partial x_j^\theta}}\, &\bigg)^{\ell_j}\, 
W^{[a]}_{\lambda,\mu,\nu}(x)\ =\ 0\,,\cr \forall&\ 
(\ell_1,\ell_2,\ell_3)\in \Z_\ge^3,\   
\sum_{j=1}^3\ell_j=t+1\,\bigg\}\ .\cr}} Then $W^{[a]}_{\lambda,\mu,\nu}(x)$ will 
not contribute to the sum \factorW\ if $t_a$ is greater than the level $k$, 
but does contribute if $t_a\le k$. That is, $t_a$ is a {\it threshold level} 
for $W^{[a]}_{\lambda,\mu,\nu}(x)$. 

Since \taW\ only involves $x_j^\alpha$ for $\alpha=\theta$, it appears 
to violate the symmetry of the horizontal algebra $X_r$. It therefore 
must be interpreted with care - one could just drop the higher powers 
of $x^\theta$, for example, to obtain incorrect results. 
The condition should be used to restrict the 
$X_r$-invariant summands in \factorW, as we'll discuss below. 

The concept of threshold level has been useful in 
WZW fusion, when the fusion is 
viewed as a truncation of the tensor product of simple Lie algebras 
\refs{\cmw\kmsw\bmw{--}\bkmw}. 
The decomposition of a tensor product 
$L(\lambda)\otimes L(\mu)$ of such representations can be written as
\eqn\tp{L(\lambda)\otimes L(\mu)\ =\ \bigoplus_{\nu\in P_\ge}\, 
T_{\lambda,\mu}^{\ \ \nu}\, L(\nu)\ ,}
where 
\eqn\Pnok{P_\ge\ =\ \big\{\sigma =\sum_{j=1}^r\,\sigma_j\Lambda^j\,|\, 
\sigma_j\in \Z_\ge,\,\forall j=1,\ldots,r\,\big\}\ }
(compare to \Pk).  
More important for three-point functions is 
\eqn\trip{L(\lambda)\otimes L(\mu)\otimes L(\nu)\ 
\supset T_{\lambda,\mu,\nu}\, L(0)\ ,}
where  
the triple product multiplicities 
$T_{\lambda,\mu,\nu}$ 
are 
related to the conventional 
tensor product multiplicities $T_{\lambda,\mu}^{\ \ \nu}$ by 
\eqn\Tud{T_{\lambda,\mu,\nu}\ =\ T_{\lambda,\mu}^{\ \ \nu^+}\ ,}
$\nu^+$ being the weight conjugate to $\nu$. Fusion products 
$\otimes_k$ can be written in a similar way:
\eqn\fpk{L(\lambda)\otimes_k L(\mu)\ =\ \bigoplus_{\nu\in P^k_\ge}\, 
T_{\lambda,\mu}^{(k)\ \nu}\, L(\nu)}
and 
\eqn\ftrip{L(\lambda)\otimes_k L(\mu)\otimes_k L(\nu)\ 
\supset T^{(k)}_{\lambda,\mu,\nu}\, L(0)\ ,}
where 
\eqn\fud{T^{(k)}_{\lambda,\mu,\nu}\ =\ T_{\lambda,\mu}^{(k)\ \nu^+}\ .}
That fusion is a truncated tensor product,
\eqn\ftrunc{T^{(k)}_{\lambda,\mu,\nu}\ \leq\ T^{(k+1)}_{\lambda,\mu,\nu}\ ,
\ \ \ \ \lim_{k\rightarrow\infty}\, T^{(k)}_{\lambda,\mu,\nu}\ =\ 
T_{\lambda,\mu,\nu}\ ,}
 follows from the 
Gepner-Witten depth rule \refs{\gepwit{,}\kmsw}. 

One can encode all this in a simple fashion using the threshold level. The 
triple product multiplicities can be written as a sum over terms of specific 
threshold level:
\eqn\Tsumt{T_{\lambda,\mu,\nu}\ =\ \sum_{t=0}^\infty\, 
n^{[t]}_{\lambda,\mu,\nu}\ ,}
where the threshold  multiplicity 
$n^{[t]}_{\lambda,\mu,\nu}$ is the number of couplings of threshold level 
$t$. We can write the fusion triple product multiplicity in a similar way:
\eqn\Tsumt{T^{(k)}_{\lambda,\mu,\nu}\ =\ \sum_{t=0}^k\, 
n^{[t]}_{\lambda,\mu,\nu}\ .} 
To compare this with a similar result for three-point functions, we define 
$W^{[a]}_{\lambda,\mu,\nu}(x;t)\, :=\, W^{[a]}_{\lambda,\mu,\nu}(x)$, if 
$t_a=t$. The goal is to write 
\eqn\factorWt{F_{\lambda,\mu,\nu}^{(k)}(x)\ 
=\ \sum_{t=0}^k\ \sum_{a=1}^{n^{[t]}_{\lambda,\mu,\nu}}\ 
W^{[a]}_{\lambda,\mu,\nu}(x;t)\ \ ,}
where the values of the index $a$ have been redefined appropriately. 

We emphasise here that a choice must be made to write \factorWt. Due to the 
syzygies, 
different representations of $W^{[a]}$ will 
exist, in general with different threshold levels 
$t_a$.

By \taW, one sees that the threshold level of a product of polynomials is 
just the sum of their individual threshold levels. We define the 
threshold level $t(E)$ of an elementary coupling $E\in {\cal E}$ by 
\eqn\tE{t(E)\ :=\ \min\,\bigg\{\, t\in \Z_\ge\ |\  \prod_{j=1}^3\, 
\bigg(\,{\partial\over{\partial x_j^\theta}}\, \bigg)^{\ell_j}\, 
R^E(x_1,x_2,x_3)\ =\ 0\,,\ \forall 
\sum_{j=1}^3\ell_j=t+1\,\bigg\}\ .} 
Then when \WR\ gets updated 
to 
\eqn\WRt{W^{[a]}_{\lambda,\mu,\nu}(x;t_a)\ =
\ \prod_{E\in{\cal E}}\ 
\big[\,R^E(x)\,\big]^{p_a(E)}\ , }
we must have 
\eqn\tsum{t_a\  =\ \sum_{E\in{\cal E}}\, p_a(E)\, t(E)\  .}

For $X_r=A_r\cong su(N)$ (so that $N=r+1$), the couplings counted in 
$T_{\lambda,\mu,\nu}$ are in one-to-one correspondence with 
Berenstein-Zelevinsky (BZ) triangles 
\ref\bz{A.D. Berenstein, A.V. Zelevinsky, J. Alg. Comb. {\bf 1} (1992) 7} 
of weight $\{\lambda,\mu,\nu\}$. The BZ triangles were used to study 
$su(3)$ fusion in \refs{\kmsw{,}\bmw}; $su(4)$ fusion was studied 
in \bkmw, which also alluded to $su(N)$ generalisation.   
For $su(3)$ and $su(4)$, it was found that every BZ triangle 
could be assigned  a threshold level. We will now ``lift'' those results 
to the corresponding polynomials, by associating a BZ triangle 
to each polynomial $W^{[a]}(x;t)$. Not only will this exercise 
promote the previous results to something containing more information, 
it will also provide the justification for some of them. 

\newsec{$su(3)$ three-point functions, triangles and fusion}

An $su(3)$ BZ triangle of weight $\{\lambda,\mu,\nu\}$ looks like:
\eqn\trithree{
 \matrix{\quad\cr m_{13}\cr
	n_{12}~~\quad l_{23}\cr
 m_{23}~\quad\qquad ~~m_{12}\cr
 n_{13}~\quad l_{12} \qquad n_{23} \quad~ l_{13} \cr \quad\cr}
\ \ .}
Its entries $l_{ij},m_{ij},n_{ij}\in \Z_\ge$ determine the Dynkin labels 
of $\lambda,\mu,\nu\in P_\ge$: 
\eqn\outthree
{ \matrix{\eqalign{m_{13}+n_{12}&=\lambda_1\ ,
  \cr m_{23}+n_{13}&=\lambda_2\ ,\cr}
 \eqalign{\ \ \ \ n_{13}+l_{12}&=\mu_1\ ,\cr n_{23}+l_{13}&=\mu_2\ ,\cr}
 \eqalign{\ \ \ \ l_{13}+m_{12}&=\nu_1\ ,\cr l_{23}+m_{13}&=\nu_2\ .\cr}\cr }
}
We call these the outer constraints on the BZ entries. 

For general $su(N)$, the BZ entries determine three elements of 
$\Z_\ge R_>$:
\eqn\lmn{\eqalign{\lambda+\mu-\nu^+\ &=\ 
\sum_{\alpha\in R_>}\, n_\alpha\,\alpha 
\ =:\ n\cdot R_>\cr
\nu+\lambda-\mu^+\ &=\ \sum_{\alpha\in R_>}\, m_\alpha\,\alpha 
\ =:\ m\cdot R_>\cr
\mu+\nu-\lambda^+\ &=\ \sum_{\alpha\in R_>}\, l_\alpha\,\alpha 
\ =:\ l\cdot R_>\ .\cr}} 
Using an orthonormal basis $\{\, e_j\, |\, j=1,\ldots,N;\ e_i\cdot e_j = 
\delta_{i,j}\, \}$ of $\R^N$, the positive roots can be written as 
\eqn\Rpos{R_>\ =\ \big\{\, e_i-e_j\ |\ 1\le i<j\le N\ \big\}\ .} 
This explains the indices $ij$ on the BZ entries. 

If the $su(3)$ version of \lmn\ is written, by applying the 
conjugation operation ${}^+$, one can derive the consistency 
conditions 
\eqn\hexthree{
\eqalign{n_{12}+m_{23}&=n_{23}+m_{12}\ ,\cr l_{12}+m_{23}&=
 l_{23}+m_{12}\ ,\cr l_{12}+n_{23}&=l_{23}+n_{12}\ .\cr}
}
These are the hexagon constraints, so named because of the geometry of 
\trithree. 

The triple tensor product multiplicity $T_{\lambda,\mu,\nu}$ 
equals the number of BZ triangles with non-negative integer entries that 
satisfy both the outer and hexagon constraints. Now a sum of 
BZ triangles is also a BZ triangle, but with a new weight equalling 
the sum of the weights of the added triangles. Compare this to \wtsum; 
a product of ``three-point polynomials'' ($R^E$'s, and so by extension 
$W_{\lambda,\mu,\nu}^{[a]}$'s) is also a three-point polynomial, 
but with the weights added. Thus, adding triangles corresponds to multiplying 
polynomials. Clearly then, we should find the triangles associated to the 
elementary polynomials, the elementary triangles. Then the 
triangle associated with an arbitrary polynomial 
$W_{\lambda,\mu,\nu}^{[a]}(x)$ can be 
found simply from its factorisation into powers of elementary 
polynomials. 

The elementary polynomials for $su(3)$ and $su(4)$ were worked out 
completely in \jrthree. Recall that $\alpha_b$ denotes the $b$-th simple root. 
For $su(3)$, $R_> = \{\,\alpha_1, 
\theta=\alpha_1+\alpha_2, \alpha_2\, \}$. We write $x_j^{\alpha_b} 
=: x_j^b$ for the flag coordinates related to the simple roots. The 
variable $x^\theta$ only appears in the combinations
\eqn\xpm{x_j^\pm\ :=\ {1\over 2}x_j^1x_j^2\, \pm\, x_j^\theta\ .}
This can be seen  as follows. Let $F_\alpha$ denote the lowering operator of 
$X_r$ associated with $\alpha\in R_>$. In the differential operator 
realisation of the algebra $X_r$, one finds \jrfree\ 
\eqn\Fop{F_\alpha(x,\partial,\lambda)\ =\ \sum_{\beta\in R_>}\, 
V^\beta_{-\alpha}(x)\, 
{\partial\over{\partial x^\beta}}\ +\ \sum^r_{j=1}\, P^j_{-\alpha}
(x)\, 
\lambda_j\ .} 
Here $V^\beta_{-\alpha}(x)$ and $P^j_{-\alpha}(x)$ denote polynomials, 
explicitly written in \jrfree, and only involving $x^\theta$ in the 
combinations \xpm. This property is carried over to the elementary 
polynomials, by a straightforward analysis.

Although for general $X_r$ it does not hold, for $su(3)$ and $su(4)$, an 
elementary coupling $E\in {\cal E}$ is uniquely specified by its weight. 
We will label the corresponding polynomials with those weights. $su(3)$ has 
8 elementary polynomials. Six are related to  
\eqn\Ri{R^{1,2,0}(x)\ :=\ R^{\Lambda^1,\Lambda^2,0}(x_1,x_2,x_3)\ =\ 
x^+_1+x_2^--x_1^1x_2^2\ \ ;}
they can be obtained from \Ri\ by permuting the subscripts of $x_1,x_2,x_3$ 
and the three weights $\Lambda^1,\Lambda^2,0$ in the identical way. 
These six elementary polynomials are actually elementary polynomials 
for two-point couplings - see \ref\jrtwo{J. Rasmussen, Mod. Phys. Lett. 
{\bf A13} (1998) 1213; 1281}. The other 
two elementary polynomials are 
\eqn\Riii{R^{1,1,1}(x)\ :=\ R^{\Lambda^1,\Lambda^1,\Lambda^1}(x_1,x_2,x_3)\ 
=\ x_2^1x_3^+-x_2^+x_3^1+x_1^1(x_2^+-x_3^+)+x_1^+(x_3^1-x_2^1)\ ,}
and $R^{2,2,2}(x) := R^{\Lambda^2,\Lambda^2,\Lambda^2}(x_1,x_2,x_3)$, 
obtained from $R^{1,1,1}(x)$ by $x_j^+\rightarrow x_j^-$ and 
$x_j^1 \rightarrow x_j^2$. 

The elementary $su(3)$ BZ triangles were written in \kmsw. They are 
\eqn\etriii{\matrix{ 
    \matrix{\quad\cr 0\cr
	1~~\quad 0\cr
 0~\quad\qquad ~~0\cr
 0~\quad 0 \qquad 1 \quad~ 0 \cr \quad\cr}
&
\qquad
&
    \matrix{\quad\cr 0\cr
	0~~\quad 0\cr
 0~\quad\qquad ~~0\cr
 1~\quad 0 \qquad 0 \quad~ 0 \cr \quad\cr}
&
\qquad
&
    \matrix{\quad\cr 0\cr
	1~~\quad 0\cr
 0~\quad\qquad ~~1\cr
 0~\quad 1 \qquad 0 \quad~ 0 \cr \quad\cr}
\cr}
}
and those obtained by applying the triangle symmetries that 
generate the dihedral group $D_3\cong S_3$. The $S_3$ orbits of the 
first two triangles of \etriii\ are 3-dimensional, and that of the last is 
2-dimensional, giving the correct number of 8 elementary triangles. 

It is simple to assign these elementary triangles to elementary polynomials. 
For example, those of \etriii\ correspond to $R^{1,2,0}$, $R^{2,1,0}$ and 
$R^{1,1,1}$, respectively. The threshold levels are also easily found. 
Because $x^\theta$ only appears in the combinations $x^\pm$ of \xpm, 
the threshold level is simply the maximum power of $x^\pm$. By \Ri\ and \Riii, 
we see that all 8 elementary polynomials (and so elementary couplings $E$) 
have threshold level $t(E)=1$.  

If we write $\lambda = \lambda_1\Lambda^1 
+\lambda_2\Lambda^2 = (\lambda_1,\lambda_2)$, then consider the 
example of the three-point 
function with weight $\{(1,1),(1,1),(1,1)\}$. The 
tensor product multiplicity is 2, with threshold levels 2 and 3 for 
the two couplings. We write 
\eqn\Wiia{W_{(1,1),(1,1),(1,1)}^{[1]}(x;2)\ =\ R^{1,1,1}(x)\, R^{2,2,2}(x)\ }
and 
\eqn\Wiib{W_{(1,1),(1,1),(1,1)}^{[2]}(x;3)\ =\ R^{2,1,0}(x) R^{0,2,1}(x) 
R^{1,0,2}(x)\ ,}
for the polynomials of threshold levels 2 and 3, respectively. 

But another product $R^{1,2,0}R^{0,1,2}R^{2,0,1}$ has weight 
$\{(1,1),(1,1),(1,1)\}$ - why is it not included? For $su(3)$, there is a 
single syzygy 
\eqn\syz{R^{1,1,1}R^{2,2,2}\ =\ R^{1,2,0}R^{0,1,2}R^{2,0,1}\,+\, 
R^{2,1,0}R^{0,2,1}R^{1,0,2}\ } 
to be taken into account. Only two linear combinations 
of the three $R$-monomials in \syz\ 
can be included, since three are not independent. 
Here the triangle correspondence may guide us, since 
there are no relations on the 
BZ triangles themselves - they are simply counted 
to get the tensor product multiplicity. 
Associating a BZ triangle to 
each elementary polynomial implements the syzygies automatically: 
both $R^{1,1,1}R^{2,2,2}$ and $R^{1,2,0}R^{0,1,2}R^{2,0,1}$ 
correspond to the same BZ triangle. 
In order to produce the threshold levels $\{2,3\}$, it is  
the product $R^{1,2,0}R^{0,1,2}R^{2,0,1}$ that should be excluded. 

That means that if 
\eqn\WRp{\eqalign{W_{\lambda,\mu,\nu}^{[a]}\ =\ 
\big(R^{1,2,0}\big)^{p_{1,2,0}}\, 
\big(R^{0,1,2}\big)^{p_{0,1,2}}\,&
\big(R^{2,0,1}\big)^{p_{2,0,1}}\,\cr
\times\,\big(R^{2,1,0}\big)^{p_{2,1,0}}\,
\big(R^{0,2,1}\big)^{p_{0,2,1}}\,
\big(R^{1,0,2}\big)^{p_{1,0,2}}\,&
\big(R^{1,1,1}\big)^{p_{1,1,1}}\,
\big(R^{2,2,2}\big)^{p_{2,2,2}}\,
\ ,\cr }}
then at least one of $p_{1,2,0},p_{0,1,2},p_{2,0,1}$ must vanish. Now,  
\WRp\ corresponds to the triangle 
\eqn\bzRiii{
 \matrix{\quad\cr p_{1,0,2}\cr\ \cr
	p_{1,2,0}+p_{1,1,1}~~\quad p_{0,1,2}+p_{2,2,2}\cr\ \cr
 p_{2,0,1}+p_{2,2,2}~\quad\qquad ~~p_{2,0,1}+p_{1,1,1}\cr\ \cr
 p_{2,1,0}~\quad p_{0,1,2}+p_{1,1,1} \qquad p_{1,2,0}+p_{2,2,2} \quad~ 
p_{0,2,1} \cr \ \cr}
\ .}
Let's assume that $p_{1,2,0}=0$. Because $t(E)=1$ for all 
elementary couplings, we see that the threshold level of \WRp\ is just 
the sum of all the powers $p_{i,j,k}$, or $t \,=\,  
n_{13}+m_{13}+l_{23}+m_{12}+l_{13}$ $=\, n_{13}+\nu_1+\nu_2$. Notice that 
$p_{1,2,0}=0$ ensures that $n_{13}+\nu_1+\nu_2 \ge 
\max\{\, m_{13}+\mu_1+\mu_2, l_{13}+\lambda_1+\lambda_2\,\}$. By 
considering the other possibilities, $p_{0,1,2}=0$ and $p_{2,0,1}=0$,  
we find the cyclically symmetric formula    
\eqn\tbziii{t\ =\ \max\big\{\, l_{13}+\lambda_1+\lambda_2, 
m_{13}+\mu_1+\mu_2, 
n_{13}+\nu_1+\nu_2\, \big\}\ ,}
confirming the result of \kmsw.

Of course, our results constitute much more than a verification of 
the results of \kmsw\ on fusion multiplicities. 
We have shown how to construct the full three-point functions \WRp. 
In addition, we have  
demonstrated that the fusion results ``lift'' in a straightforward manner 
to those for three-point functions. 

Furthermore, in \kmsw\ certain obstacles were found in 
a direct application of the Gepner-Witten 
depth rule. Here we have shown how to implement the depth rule
in a very simple way, using a  
differential operator realisation of $X_r$. We will have more to say on 
this subject below. 

The triangles helped us implement the 
syzygy \syz, by choosing a product to forbid, 
$R^{1,2,0}R^{0,1,2}R^{2,0,1}$. But a different choice could 
certainly have been made, with no effect on the results at large level. 
Because the syzygy is inhomogeneous in the threshold level, 
however, the choice becomes important at smaller levels. For example, 
if we had forbidden the product $R^{1,1,1}R^{2,2,2}$, then 
the additivity \tsum\ of threshold levels would be lost. Another 
possibility is to forbid  $R^{2,1,0}R^{0,2,1}R^{1,0,2}$. This is 
equivalent to choosing the orientation around the BZ triangle that is 
opposite to that used in \trithree. Of course, one could also forbid a 
linear combination of the $R$-monomials in \syz, but such a choice 
is not convenient for us.     

It is often preferable to work directly with the polynomials 
$W^{[a]}_{\lambda,\mu,\nu}(x)$, rather than to first express them 
as products of the elementary polynomials $R^E(x)$. But we can find the 
triangle of a polynomial $W^{[a]}_{\lambda,\mu,\nu}(x)$ directly. 
For our purposes, rewrite \lmn\ as
\eqn\newlmn{\eqalign{\lambda+
\bigg(\mu-\sum_{\alpha\in R_>}\, n_\alpha\,\alpha\bigg)+(-\nu^+)\ &=\ 0 
\cr
\nu+
\bigg(\lambda-\sum_{\alpha\in R_>}\, m_\alpha\,\alpha\bigg)+(-\mu^+)\ &=\ 0 
\cr
\mu+
\bigg(\nu-\sum_{\alpha\in R_>}\, l_\alpha\,\alpha\bigg)+(-\lambda^+)\ &=\ 0\ . 
\cr}}
The first of these expressions suggests that we look at the contribution 
$|\lambda\rangle \otimes |\mu^\prime\rangle \otimes |-\nu^+\rangle$. Here 
$|\lambda\rangle$ is the highest weight state of $L(\lambda)$, 
$|-\nu^+\rangle$ is the lowest weight state of $L(\nu)$, and 
$|\mu^\prime\rangle$ denotes a state of $L(\mu)$, of weight 
$\mu^\prime:= \mu-n\cdot R_>$. 

Let 
$x_3^{[\beta]}$ denote any linear combination of products  
\eqn\xwt{\prod_{\alpha\in R_>}\,(x_3^\alpha)^{b_\alpha}}
such that $\sum b_\alpha\,\alpha = \beta$. 
By construction, , 
$x_1^{[0]}=1$ isolates the highest weight state $|\lambda\rangle$,  
while the lowest weight state 
$|-\nu^+\rangle$ corresponds to $x_3^{[\nu+\nu^+]}$ 
(see \Kpoly, for example). Now, a fixed 
$W_{\lambda,\mu,\nu}^{[a]}$ should include a unique term  
(with non-vanishing coefficient) 
$x_1^{[0]}\, x_2^{[\mu-\mu^\prime]}\, x_3^{[\nu+\nu^+]}$, or 
$x_1^{[0]}\, x_2^{[n\cdot R_>]}\, 
x_3^{[\nu+\nu^+]}$. Let that term be denoted 
$W_{\lambda,\mu,\nu}^{[a]}(x) \vert x_1^{[0]}\, x_2^{[n\cdot R_>]}\, 
x_3^{[\nu+\nu^+]}$. 
We find that its factor  
$x_2^{[n\cdot R_>]}$ suffices to determine 
the $\{n_\alpha\}$, i.e. one third of the BZ entries. The other two 
thirds are obtained in a similar way, using the other two equations 
of \newlmn. 

Simply put, we find 
\eqn\xiipr{ x_2^{[n\cdot R_>]}\ =\ 
(x_2^1)^{n_{12}}\  (x_2^-)^{n_{13}}\ (x_2^2)^{n_{23}}\  .}
This result and its two siblings can be encoded in the form of 
a BZ triangle:
\eqn\trix{
 \matrix{\quad\cr x_1^-\cr
	x_2^1~~\quad x_3^2\cr
 x_1^2~\quad\qquad ~~x_1^1\cr
 x_2^-~\quad x_3^1 \qquad x_2^2 \quad~ x_3^- \cr \quad\cr}\ \ .
}
This diagram can be used to find the BZ triangle corresponding to a 
given polynomial $W_{\lambda,\mu,\nu}^{[a]}(x)$. We should emphasise, however, 
that it does not allow us to construct the full polynomial from a 
prescribed BZ triangle, only a few terms in it.

A refined version of the Gepner-Witten depth rule \gepwit\  
was conjectured in \refs{\kmsw{,}\kmswii} 
(see also \ref\mwcjp{M.A. Walton,
Can. J. Phys. {\bf 72} (1994) 527}). It is less symmetric than \gwsym, but 
is a natural generalisation of a well-known 
formula in the theory of simple Lie algebras. It  
will prove to be more convenient in 
some computations. The realisation-independent form of the formula is
\eqn\gwns{T^{(k)}_{\lambda,\mu,\nu}\ =\ \dim\{\ v\in L(\mu; \nu^+-\lambda)
\ |\   
F_{\alpha_i}^{\ \nu^+_i+1}\, v\,= \, 0,\ 
\forall i\in\{1,2,\ldots,r\};\ 
E_\theta^{\ k-\nu\cdot\theta+1}\,v \,=\, 0\ \}\ .}
Here $L(\mu; \nu^+-\lambda)$ is the subspace of $L(\mu)$ 
of weight $\nu^+-\lambda$, and $\nu^+_i$ is the $i$-th Dynkin label of 
$\nu^+$. The well-known classical formula for tensor product 
multiplicities (see \ref\zelo{D.P. Zhelobenko, Compact Lie Groups and their 
Representations (American Mathematical Society, 1973), Theorem 4, section 78}, 
for example) 
is obtained in the limit of large level $k\rightarrow \infty$. Then the 
constraint $E_\theta^{\ k-\nu\cdot\theta+1} v$ $=0$ is always satisfied, 
and can be dropped.  

If written similar to \gwsW, it is 
\eqn\gwnsW{\eqalign{0\ =\  
\bigg(\,{\partial\over{\partial x_2^\theta}}\, \bigg)^{\ell}\, 
&W^{[a]}_{\Lambda_{(1)},\Lambda_{(2)},\Lambda_{(3)}}(x)
\vert x_1^{[0]}\, x_2^{[n\cdot R_>]}x_3^{[\nu+\nu^+]}
\ ,\cr 
&\forall\ \ell\in\Z_\ge\, ,\ 
{\rm such\ that\ }\ell\, >\, k\,-\,\nu\cdot\theta\ .}}
Consequently, we find 
\eqn\taWns{t_a\ =\ \min\bigg\{\, t\in \Z_\ge\ |\  
\bigg(\,{\partial\over{\partial x_2^\theta}}\, \bigg)^{t+1}\, 
W^{[a]}_{\lambda,\mu,\nu}(x) 
\vert x_1^{[0]}\, x_2^{[n\cdot R_>]}\, 
x_3^{[\nu+\nu^+]}\ =\ 0\ \bigg\}\ +\ \nu\cdot\theta\ .} 
Clearly, the last two formulas treat the 3 weights in asymmetric manner.  
Consequently, 
5 other formulas of this type can be written, based on 
the 5 other permutations of $\lambda,\mu,\nu$. 

Using these formulas, we easily verify the threshold levels computed 
using the symmetric version of the depth rule. 
This provides evidence for the refinement of the depth 
rule \gwns, conjectured in \refs{\kmsw{,}\kmswii}. This asymmetric form 
has the advantage that it effectively reduces consideration from 
the coupling of three states to a single state, one that necessarily 
couples to a simple highest-weight state and a simple lowest-weight state. 
Consequently, to use \gwnsW, we need only a small part of the 
polynomial $W^{[a]}_{\lambda,\mu,\nu}$; for  \gwsW, the full 
polynomial is required.

\newsec{$su(4)$ three-point functions, triangles and fusion}

The results for $su(4)$ are technically more complicated than those for 
$su(3)$. They also include an interesting new feature, discussed below.  

The 18 elementary couplings may be  
specified by their weights, $\{\lambda,\mu,\nu\}$, or $\{\Lambda_{(1)}, 
\Lambda_{(2)}, \Lambda_{(3)}\}$  
\jrthree. Permuting 1,2,3 in 
$\{\Lambda_{(1)}, 
\Lambda_{(2)}, \Lambda_{(3)}\}$ 
and  $x_1,x_2,x_3$ produces other elementary couplings. The 
polynomial-triangle correspondence breaks the permutation $S_3$ symmetry 
to the cyclic $\Z_3$ symmetry (see \newlmn, e.g.). So only 
cyclic permutations of 1,2,3 will produce other elementary couplings with 
related BZ triangles.

Of the 18 elementary couplings, 9 are of the 
two-point type \jrtwo. They separate into three cyclic-permutation orbits,  
with representative polynomials   
\eqn\Rivii{R^{1,3,0}:= R^{\Lambda^1,\Lambda^3, 0}\ ,\ \ 
R^{3,1,0}:= R^{\Lambda^3,\Lambda^1, 0} 
\ \ {\rm and}\ \ 
R^{2,2,0}:= R^{\Lambda^2,\Lambda^2, 0}\ \ .}
The 9 elementary polynomials of three-point type 
form another three orbits, with representatives
\eqn\Riviii{R^{1,1,2}:= R^{\Lambda^1,\Lambda^1, \Lambda^2},\  
R^{2,3,3}:= R^{\Lambda^2,\Lambda^3, \Lambda^3},\ {\rm and}\  
R^{13,2,2}:= R^{\Lambda^1+\Lambda^3, \Lambda^2, \Lambda^2}\ .}
By \taW\ or \taWns, we find that all these couplings have threshold level 1, 
except $R^{13,2,2}, R^{2,13,2}, R^{2,2,13}$, which have $t=2$.  

For $su(4)$ the BZ triangle is 
\eqn\bziv{
 \matrix{m_{14}\cr
	n_{12}~~\quad l_{34}\cr
 m_{24}~\qquad\qquad ~~m_{13}\cr
 n_{13}\qquad l_{23}\qquad n_{23} \qquad l_{24}\cr
 m_{34}\qquad\qquad\quad m_{23}\qquad\qquad\quad m_{12} \cr
 n_{14}\qquad~~ l_{12}\qquad~ n_{24}\quad~\quad l_{13}\quad~~~ n_{34}\qquad
 l_{14} \cr \cr }\ \ ,}
with 
corresponding Dynkin labels given by the outer constraints
\eqn\outiv{
 \matrix{\eqalign{m_{14}+n_{12}&=\lambda_1\ ,\cr m_{24}+n_{13}&=\lambda_2\ ,\cr
  m_{34}+n_{14}&=\lambda_3\ ,~~\cr}
 \eqalign{\ \ n_{14}+l_{12}&=\mu_1\ ,\cr
 n_{24}+l_{13}&=\mu_2\ ,\cr n_{34}+l_{14}&=\mu_3\ ,~~\cr}
 \eqalign{l_{14}+m_{12}&=\nu_1\ ,\cr
 l_{24}+m_{13}&=\nu_2\ ,\cr ~~l_{34}+m_{14}&=\nu_3\ .
 \cr } \cr}}
The $su(4)$ BZ triangle contains three hexagons, and there are inner 
constraints (hexagon identities) related to each of them in the obvious way.   

The elementary polynomials of \Rivii\ and \Riviii\ can be 
related easily to elementary BZ triangles. 
The non-zero BZ entries corresponding to \Rivii\ are 
\eqn\tivii{
\{\, n_{12}=n_{23}=n_{34}=1 \,\}\ ,\ \ \{\, n_{14}=1\,\}\ \ {\rm and}\ \ 
\{\, n_{13}=n_{24}=1 \,\}\ \ ;
}
and those related to 
\Riviii\ are 
\eqn\tiviii{\eqalign{
\{\, n_{12}=l_{12}=&m_{13}=l_{23}=1 \,\}\, ,\ \ 
\{\, m_{24}=l_{34}=n_{23}=n_{34}=1 \,\}\, \cr
&\ {\rm and}\ 
\{\, n_{12}=m_{34}=l_{23}=n_{24}=m_{13}=1 \,\}\ .
}} 

The $su(4)$ polynomial $\rightarrow$ triangle correspondence 
cannot be given as simply as for $su(3)$, in 
\xiipr,\trix. Instead, we write
\eqn\funi{\eqalign{K_{1,0,0}(x_2)\ \rightarrow&\ \{\}\cr
K_{-1,1,0}(x_2)\ \rightarrow&\ \{n_{12}=1\}\cr
K_{0,-1,1}(x_2)\ \rightarrow&\ \{n_{13}=1\}\cr
K_{0,0,-1}(x_2)\ \rightarrow&\ \{n_{14}=1\}\cr
}}

\eqn\funii{\eqalign{K_{0,1,0}(x_2)\ \rightarrow&\ \{\}\cr
K_{1,-1,1}(x_2)\ \rightarrow&\ \{n_{23}=1\}\cr
K_{-1,0,1}(x_2)\ \rightarrow&\ \{n_{12}=n_{23}=1\}\cr
K_{1,0,-1}(x_2)\ \rightarrow&\ \{n_{24}=1\}\cr
K_{-1,1,-1}(x_2)\ \rightarrow&\ \{n_{12}=n_{24}=1\}\cr
K_{0,-1,0}(x_2)\ \rightarrow&\ \{n_{13}=n_{24}=1\}\cr
}}

\eqn\funiii{\eqalign{K_{0,0,1}(x_2)\ \rightarrow&\ \{\}\cr
K_{0,1,-1}(x_2)\ \rightarrow&\ \{n_{34}=1\}\cr
K_{1,-1,0}(x_2)\ \rightarrow&\ \{n_{34}=n_{23}=1\}\cr
K_{-1,0,0}(x_2)\ \rightarrow&\ \{n_{34}=n_{12}=n_{23}=1\}\cr
}}

\eqn\Lnij{-L_1(x_2)+2L_2(x_2)-L_3(x_2)\ \rightarrow\ \{n_{13}=n_{34}=1\}\ }
The $K$- and $L$-polynomials were defined in \jrthree. Those in 
\funi, \funii, \funiii\  are 
in one-to-one correspondence with the states in  
$L(\Lambda^1)$, $L(\Lambda^2)$, $L(\Lambda^3)$, respectively. 
Eqn. \Lnij\ picks out one state in the 
three-dimensional subspace of weight 0 in the adjoint representation 
$L(\Lambda^1+\Lambda^3)$. 
 
The rules \funi,\funii,\funiii\ and \Lnij, can be used to find the triangle 
that corresponds to a given polynomial directly, without first 
decomposing the polynomial into its elementary factors. 

For $su(4)$, there are 15 syzygies. They can be obtained 
from the following 5, by permuting 
the superscripts cyclically in all polynomials: 
\eqn\syzi{{\cal S}\{\, R^{13,2,2}R^{2,13,2}, R^{2,2,0}R^{1,1,2}R^{3,3,2}, 
R^{0,2,2}R^{2,0,2}R^{3,1,0}R^{1,3,0} \,\}\ ,}
\eqn\syzii{{\cal S}\{\, R^{13,2,2}R^{2,3,3}, R^{1,0,3}R^{2,2,0}R^{3,3,2}, 
R^{1,3,0}R^{2,0,2}R^{3,2,3} \,\}\ ,}
\eqn\syziii{{\cal S}\{\, R^{2,2,13}R^{1,1,2}, R^{1,0,3}R^{0,2,2}R^{2,1,1}, 
R^{0,1,3}R^{2,0,2}R^{1,2,1} \,\}\ ,} 
\eqn\syziv{{\cal S}\{\, R^{13,2,2}R^{0,1,3}, R^{1,1,2}R^{3,2,3}, 
R^{1,0,3}R^{3,1,0}R^{0,2,2} \,\}\ ,}
\eqn\syzv{{\cal S}\{\, R^{2,13,2}R^{1,0,3}, R^{1,1,2}R^{2,3,3}, 
R^{0,1,3}R^{1,3,0}R^{2,0,2} \,\}\ .}
Here ${\cal S}\{A,B,C\}$ indicates there exists one independent vanishing 
linear combination 
$\alpha A+\beta B+\gamma C = 0$, with 
$\alpha,\beta,\gamma \not=0$.\foot{The explicit linear 
combinations are given in 
Ref. \jrthree. We take this opportunity to correct a sign error there: 
the parameter $u_5=-2$.} 

We will implement each  syzygy by choosing a forbidden product 
from among its ``component'' $R$-monomials.  That is, we will not 
consider eliminating linear combinations of the component $R$-monomials 
as redundant, although that is certainly possible.  
Even with this limitation, however, the choice is not unique.   
As we did for $su(3)$, we will use the triangle correspondence, 
and 
specific fusion multiplicities to guide us. We will preserve the 
additivity of the threshold level, and  by choosing a symmetric 
set of forbidden couplings,  
the cyclic $\Z_3$ symmetry of the triangles. 

The fusions relevant to 
\syzi,\syzii,\syziii,\syziv,\syzv, are
\eqn\fusi{L(1,1,1)\otimes L(1,1,1)\otimes L(0,2,0)\ \supset\ 
L(0,0,0)_{3,4}\ ,} 
\eqn\fusii{L(1,1,1)\otimes L(0,1,1)\otimes L(0,1,1)\ \supset\ 
L(0,0,0)_{3,3}\ ,} 
\eqn\fusiii{L(1,1,0)\otimes L(1,1,0)\otimes L(1,1,1)\ \supset\ 
L(0,0,0)_{3,3}\ ,} 
\eqn\fusiv{L(1,0,1)\otimes L(1,1,0)\otimes L(0,1,1)\ \supset\ 
L(0,0,0)_{2,3}\ ,} 
\eqn\fusv{L(1,1,0)\otimes L(1,0,1)\otimes L(0,1,1)\ \supset\ 
L(0,0,0)_{2,3}\ ,} 
respectively. Here the notation is $L(\lambda_1,\lambda_2, \lambda_3) 
=L(\lambda)$, with $\lambda=\sum_{i=1}^3\lambda_i\Lambda^i$. 
Also, in \fusi, e.g., $L(0,0,0)_{3,4}$ indicates that 
$T_{(1,1,1),(1,1,1),(0,2,0)}=2$, and that the corresponding set of 
threshold levels is $\{3,4\}$. Compare these threshold levels with 
those of \syzi: $\{4,3,4\}$. Either $R^{13,2,2}R^{2,13,2}$ or 
$R^{0,2,2}R^{2,0,2}R^{3,1,0}R^{1,3,0}$ must be forbidden 
(recall that we do not consider the linear combinations thereof as possible 
forbidden terms). Since 
$R^{13,2,2}R^{2,13,2}$ and $R^{2,2,0}R^{1,1,2}R^{3,3,2}$ correspond to 
the same BZ triangle, however, we forbid $R^{13,2,2}R^{2,13,2}$. 

Analysing the syzygies \syzi,\syzii,\syziii,\syziv,\syziv,   
yields the forbidden products
\eqn\fpi{R^{13,2,2}R^{2,13,2}\ ;}
\eqn\fpii{R^{13,2,2}R^{2,3,3}\ \ {\rm or}\ \  
R^{1,3,0}R^{2,0,2}R^{3,2,3}\ \ ;}
\eqn\fpiii{R^{2,2,13}R^{1,1,2}\ \ {\rm or}\ \  
R^{0,1,3}R^{2,0,2}R^{1,2,1}\ \ ;}
\eqn\fpiv{R^{13,2,2}R^{0,1,3}\ ;}
\eqn\fpv{R^{0,1,3}R^{1,3,0}R^{2,0,2}\ ,} 
respectively. The consequences of the other 10 syzygies are 
obtained by cyclically permuting the superscripts in \fpi-\fpv.  
We will specify a choice of forbidden products later.

First we treat a new feature that arises in the $su(4)$ case. When 
$\lambda=\mu=\nu=\Lambda^1+\Lambda^3$, the highest weight of the 
adjoint representation, the  tensor product multiplicity is 2, and 
the threshold levels are 2 and 3. But the possible $R$-monomials of 
this weight are 
\eqn\feat{R^{1,3,0}R^{0,1,3}R^{3,0,1}\ \ {\rm and}\ \ 
R^{3,1,0}R^{0,3,1}R^{1,0,3}\ ,}
both with threshold level 3. This discrepancy 
was noticed first in \bkmw, where 
considerations were based entirely on BZ triangles, and was 
verified in 
\ref\bcmii{L. B\'egin, C. Cummins, P. Mathieu, J. Math. Phys. {\bf 41} 
(2000) 7640}. Here we can explain 
why it happens: the terms in the $R$-monomials \feat\ 
of maximum degree in $x^\theta$ are cancelled in their sum.  
That is, 
\eqn\rnew{R^{13,13,13}\ :=\ 
R^{1,3,0}R^{0,1,3}R^{3,0,1} + R^{3,1,0}R^{0,3,1}R^{1,0,3} }
has threshold level 2, as found by \taW. 

One can persist in using the original 
set of elementary couplings obtained from 
\Rivii,\Riviii. When considering three-point functions of arbitrary weight 
$\{\lambda,\mu,\nu\}$, then linear combinations such as \rnew\ may have to be 
found, in order to be consistent with the fusion multiplicities. 
In general, however, this is a tedious task. 

Alternatively, as suggested by \rnew, we can continue to work 
in the same framework by introducing 
$R^{13,13,13}$ as a new elementary polynomial. This follows the procedure of 
\bkmw\ and \bcmii. The price to be paid is the introduction 
of extra syzygies involving the 
new elementary coupling, so that a single 
coupling is not counted more than once. In particular, we must 
implement 
\eqn\newsyz{{\cal S}\{ R^{13,13,13}, R^{1,3,0}R^{0,1,3}R^{3,0,1}, 
R^{3,1,0}R^{0,3,1}R^{1,0,3}\} }
as a new syzygy, by forbidding either $R^{1,3,0}R^{0,1,3}R^{3,0,1}$ or 
$R^{3,1,0}R^{0,3,1}R^{1,0,3}$ (for example). 
Since $R^{3,1,0}R^{0,3,1}R^{1,0,3}$ 
corresponds to the simpler triangle (with non-vanishing entries 
$l_{14}=m_{14}=n_{14}=1$) we choose to forbid it, so that the 
new coupling then corresponds to this simple triangle.   
 
This introduction of a new elementary coupling is perhaps not too 
surprising, if one looks back at the $su(3)$ case. Recall the single syzygy 
\syz\ for $su(3)$. As far as tensor products are concerned, it can be 
implemented by forbidding the product $R^{1,1,1}R^{2,2,2}$. As noted above, 
however, two other possibilities (forbidding $R^{1,2,0}R^{0,1,2}R^{2,0,1}$ or 
forbidding $R^{2,1,0}R^{0,2,1}R^{1,0,2}$) 
are more convenient for fusion, since 
they preserve in a simple way the additivity of the threshold level. 
If we insist on forbidding  $R^{1,1,1}R^{2,2,2}$, however, the situation is 
completely analogous to the $su(4)$ case just considered. The 
$R$-monomials for the coupling $\lambda=\mu=\nu=\Lambda^1+\Lambda^2$ are 
then $R^{1,2,0}R^{0,1,2}R^{2,0,1}$ and $R^{2,1,0}R^{0,2,1}R^{1,0,2}$, each 
with threshold level 3; the correct threshold levels are 2,3, however. But 
a new elementary polynomial $R^{12,12,12}:= 
R^{1,2,0}R^{0,1,2}R^{2,0,1} +  R^{2,1,0}R^{0,2,1}R^{1,0,2}$ can be 
introduced, and it has threshold level 2. Of course, a new syzygy 
${\cal S}\{ R^{12,12,12}, R^{1,2,0}R^{0,1,2}R^{2,0,1}, 
R^{2,1,0}R^{0,2,1}R^{1,0,2}\}$ is also necessary. 

The difference between the $su(3)$ and $su(4)$ cases originates in 
the syzygies. For $su(3)$, implementing the syzygy \syz\ makes the linear 
combination $R^{1,2,0}R^{0,1,2}R^{2,0,1}$ $+R^{2,1,0}R^{0,2,1}R^{1,0,2}$ 
inaccessible. No such syzygy is present in the $su(4)$ case, however. The 
combination $R^{1,3,0}R^{0,1,3}R^{3,0,1} +$ $R^{3,1,0}R^{0,3,1}R^{1,0,3}$ 
can be included, and so generates a new elementary coupling. 

It is interesting to use the asymmetric depth rule,  \gwns\ and \gwnsW, to 
analyse these phenomena at the level of states. A single relation among 
$R$-monomials implies an  
$x_1^{[m\cdot R_>]}$ relation, an $x_2^{[n\cdot R_>]}$ relation,  and 
an $x_3^{[l\cdot R_>]}$ relation. The $su(3)$ syzygy \syz, for example, 
yields 
\eqn\syzx{x^1x^2\ =\ x^-\ +\ x^+\ }
three times, once for each $x\rightarrow x_i$, $i=1,2,3$. 
In terms of lowering operators, \syzx\ is a 
consequence of the relation
\eqn\FFF{[F_{\alpha_1}, F_{\alpha_2}]\ =\ -F_{\theta}\ .} 
It 
is identical in the three cases because of the symmetry of \syz, and encodes 
a relation between three states of weight 0 in the $su(3)$ representation 
$L(1,1)$. Since only two such states are independent, a relation 
like \syzx\ is inevitable. A basis of the two-dimensional space can be 
chosen with depths $\{1,1\}$, or with $\{1,0\}$, giving rise to 
threshold levels 
$\{3,3\}$ or $\{3,2\}$. Since the latter is correct for 
$L(1,1)^{\otimes 3} \supset L(0)$, we see that choosing the correct 
$R$-monomial $R^{1,1,1}R^{2,2,2} =$ 
$R^{1,2,0}R^{0,1,2}R^{2,0,1} + R^{2,1,0}R^{0,2,1}R^{2,0,1}$ 
corresponds to choosing a basis of states of the required depths. 

Many of the $su(4)$ syzygies are not symmetric and so lead 
through the asymmetric depth rule to 
three distinct $x_i$-relations. But the definition of the new 
elementary coupling \rnew, is symmetric. It yields 
\eqn\syzyx{x^1x^{23} + x^{12}x^3\ =\ -K_{-100}(x) + K_{00-1}(x)\ ,}
in the notation of \jrthree, where 
\eqn\KKx{\eqalign{K_{00-1}\ =&\ x^\theta\ +\ 
{1\over 2}x^1x^{23}\  +\ {1\over 2}x^{12}x^{3}\ +\ 
{1\over 6}x^1x^2x^3\ ,\cr
K_{-100}\ =&\ x^\theta\ -\  
{1\over 2}x^1x^{23}\  -\ {1\over 2}x^{12}x^{3}\ +\ 
{1\over 6}x^1x^2x^3\ .\cr}}
This indicates that a linear combination of two depth-1 states of weight 
0 in $L(1,0,1)$ is a state of depth 0. 

In an arbitrary basis, finding such linear 
combinations are necessary when using the depth rule \refs{\kmsw{,}\kmswii}. 
It is not suprising, therefore, that the new coupling \rnew\ must be 
introduced; it is perhaps surprising that \rnew\ is the only  
new coupling needed. This has not been proved, however. We defer to \bcmii, 
where some counting arguments supported this assumption. 

We conclude our construction of $su(4)$ three-point functions by listing 
a choice of a set of forbidden couplings:
\eqn\fbset{\eqalign{\{\ R^{13,2,2}R^{2,13,2},\ 
&R^{2,13,2}R^{2,2,13},\ R^{2,2,13}R^{13,2,2},\ \cr
R^{13,2,2}R^{2,3,3},\ 
&R^{2,13,2}R^{3,2,3},\ R^{2,2,13}R^{3,3,2},\ \cr 
R^{2,2,13}R^{1,1,2},\ 
&R^{13,2,2}R^{2,1,1},\ R^{2,13,2}R^{1,2,1},\ \cr 
R^{13,2,2}R^{0,1,3},\ 
&R^{2,13,2}R^{3,0,1},\ R^{2,2,13}R^{1,3,0},\ \cr 
R^{0,1,3}R^{1,3,0}R^{2,0,2},\ &R^{3,0,1}R^{0,1,3}R^{2,2,0},\ 
R^{1,3,0}R^{3,0,1}R^{0,2,2},\ \cr 
&R^{3,1,0}R^{0,3,1}R^{1,0,3}\ \}\ .\cr  
}}
The three-point function of weight $\{\lambda,\mu,\nu\}$ can be 
constructed from the sum over all $R$-monomials of that weight 
(as in \factorW,\WR) that do not contain the forbidden products \fbset.

\nref\rwfustri{J. Rasmussen, 
M.A. Walton, Affine $su(3)$ and $su(4)$ fusion multiplicities as 
polytope volumes, hep-th/0106287}

To avoid confusion, we should state that our choice of forbidden 
products will not lead to the formula of \refs{\bkmw{,}\bcmii} assigning 
a threshold level to each $su(4)$ BZ triangle. Since it is only the 
set of threshold levels for a fixed weight $\{\lambda,\mu,\nu\}$
that is physically relevant, this is not 
a problem. We have not attempted to find a different formula, 
resulting from the choice \fbset,  assigning a 
threshold level to a BZ triangle. Threshold levels can be very easily 
found from the relevant $R$-monomials  (using \tsum), and these 
$R$-monomials must be constructed in order to write the three-point 
function. When the interest is just fusion, and not the three-point 
functions, a formula like that of \refs{\bkmw{,}\bcmii} is very helpful 
(see \rwfustri, e.g.). Here it represents an intermediate step we can 
eliminate.

\newsec{Discussion}  

Building on the work \jrthree, we have shown how to 
construct WZW three-point functions for 
generating-function primary fields, for 
the algebras $su(3)$ and $su(4)$. 
More generally, our approach is ideally suited to the 
use on three-point functions of the Gepner-Witten depth 
rule, and makes it 
clear that the level-dependence of WZW three-point functions mimics that 
of the corresponding fusion multiplicities, for all algebras. 

In addition,  
we showed that a refined version  \refs{\kmsw{,}\kmswii{,}\mwcjp} 
of the original depth rule leads to a correspondence between 
three-point functions and BZ triangles. This correspondence was very 
helpful for $su(3)$ and $su(4)$.  
For example, it allowed us to use previous work on $su(3)$ 
and $su(4)$ 
fusion and BZ triangles \refs{\kmsw{-}\bkmw}. It also helped eliminate the 
redundancies encoded in the syzygies.  

We will conclude with a brief discussion of possible uses and improvements 
of this work. We hope to address some of these issues in the future. 
 
To be more precise, we found a correspondence between the polynomials 
$R^E(x)$ and $W^{[a]}_{\lambda,\mu,\nu}(x)$ and the BZ triangles, for 
the algebras $su(3)$ and $su(4)$. (For completeness, the $su(2)$ case is 
treated in a short appendix.) That is,  
given such a polynomial, 
we showed how to find the associated triangle. 
The polynomials themselves are more important than the triangles, 
however. It 
would  be interesting to try to find an algorithm to 
construct the appropriate polynomial from a BZ triangle. 

As described above, when respecting the additivity of the 
threshold level, the triangles may favour certain ways of imposing the 
syzygies. However, it is sometimes preferable to impose them in other ways. 
Furthermore, BZ triangles, or their analogues, have not been constructed 
for all algebras. If one is interested in $G_2$ three-point functions, 
for example, 
one cannot rely on BZ constructions as a guide.

\nref\rwNh{J. Rasmussen, M.A. Walton, Fusion multiplicities as 
polytope volumes: ${\cal N}$-point and 
higher-genus $su(2)$ fusion, hep-th/0104240}

In those cases, one should consider the polynomials themselves, instead of 
first projecting to the corresponding triangles, and working with them. 
For example,  
one can write polytope volume formulas for the 
fusion multiplicities 
\ref\rwinp{J. Rasmussen, M.A. Walton, work in progress} that are analogous 
to those written in \ref\msum{J. Rasmussen, M.A. Walton, $su(N)$ 
tensor product multiplicities and virtual Berenstein-Zelevinsky triangles, 
math-ph/0010051} for the tensor product multiplicities for 
$su(N)$. These multiple sum 
formulas 
generalise the expressions for 
fusion multiplicities in 
\refs{\bmw{,}\rwNh{,}\rwfustri}, 
derived using 
BZ techniques. In particular, they are derived using virtual three-point 
functions, generalisations of the virtual BZ triangles and diagrams used in 
\refs{\bmw{,}\msum{,}\rwNh{,}\rwfustri}.  

Finally, consider the generalisation to higher rank -- for $su(N)$ or other 
algebras. Our method of studying the level-dependence should work simply 
for any algebra, of any rank. However, we have relied 
on the method of elementary couplings to construct the polynomials 
$W_{\lambda,\mu,\nu}^{[a]}$, before the level-dependent constraints are 
imposed. Since the numbers of elementary couplings and syzygies climb 
very rapidly, this 
part of the construction becomes impractical for 
higher ranks. A method of finding the 
polynomials $W_{\lambda,\mu,\nu}^{[a]}$ 
directly, without first factoring into elementary polynomials, might  
make larger ranks more tractable.

\vskip 1truecm
\noindent{\it Acknowledgement}

We thank Chris Cummins and Pierre Mathieu for comments on the manuscript.

\appendix{A}{$su(2)$ three-point functions, triangles and fusion}

Since there is only one positive root, $\alpha_1=\theta$, we use 
$x_j^1=x^\theta_j=x_j$, $j=1,2,3$. Similarly, we write
\eqn\EHF{\eqalign{E_{\alpha_1}(x,\partial,\lambda) &=: E(x,\partial,\lambda)
={\partial\over{\partial x}}\ \cr
H_{\alpha_1}(x,\partial,\lambda) &=: H(x,\partial,\lambda)
= -2x{\partial\over{\partial x}}+\lambda\ \cr 
F_{\alpha_1}(x,\partial,\lambda) &=: F(x,\partial,\lambda)
= -x^2{\partial\over{\partial x}}+\lambda x\ \cr
}}
for the differential-operator realisation of $su(2)$.  

There are three elementary couplings, 
\eqn\Riio{R^{1,1,0}(x)\ =\ x_1-x_2\ ,\ \ 
R^{0,1,1}(x)\ =\ x_2-x_3\ ,\ \ 
R^{1,0,1}(x)\ =\ x_3-x_1\ ,}
with no syzygies. For a weight $\{\lambda, \mu, \nu\} = 
\{\lambda_1\Lambda^1, \mu_1\Lambda^1, \nu_1\Lambda^1\}$, we therefore find
\eqn\Wlmnii{W_{\lambda,\mu,\nu}\ =\ 
\left[R^{1,1,0}\right]^{(\lambda_1+\mu_1-\nu_1)/2}\ 
\left[R^{0,1,1}\right]^{(-\lambda_1+\mu_1+\nu_1)/2}\ 
\left[R^{1,0,1}\right]^{(\lambda_1-\mu_1+\nu_1)/2}\ .}

An $su(2)$ BZ triangle is very simple:
\eqn\tritwo{
 \matrix{\quad\cr m\cr
	n~~\quad l\cr}
\ \ \qquad.}
To be consistent with \trithree\ and \bziv, we should write $m=m_{12}$, etc., 
since $\alpha_1=e_1-e_2$, where $\{e_1,e_2\}$ is an orthonormal basis of 
$\R^2$. We dropped the subscripts for simplicity, however. The outer 
constraints are 
\eqn\outii{\lambda_1=m+n\ ,\ \ \mu_1=n+l\ ,\ \ \nu_1=l+m\ ,}
and there are no inner constraints, since the triangle contains no 
hexagons.

The elementary polynomials of \Riio\ are associated to the triangles
\eqn\Riiotri{\matrix{\quad\cr 0\cr
	1~~\quad 0\cr}\ \ \ \ ,\ \ \qquad
\matrix{\quad\cr 0\cr
	0~~\quad 1\cr}\ \ \ \  ,\ \ \qquad
\matrix{\quad\cr 1\cr
	0~~\quad 0\cr}\ \ \ \ ,\ } 
respectively. Comparing this correspondence with \Wlmnii, we see that 
\eqn\lmnla{(\lambda_1+\mu_1-\nu_1)/2=n\ ,\ \ 
(-\lambda_1+\mu_1+\nu_1)/2=l\ ,\ \ 
(\lambda_1-\mu_1+\nu_1)/2=m\ ,\ }
as must be, by the outer constraints \outii. 

The symmetric form \gwsW\ of the depth rule gives threshold level 
$t=1$ for the three 
elementary couplings, and threshold $t=l+m+n = (\lambda_1+\mu_1+\nu_1)/2$ 
for the general polynomial  
\Wlmnii. This is consistent with the additivity of the threshold 
level under multiplication of $R$-monomials. 

The asymmetric form \gwnsW\ of the depth rule is in agreement. For the general 
polynomial \Wlmnii, we find
\eqn\mnlR{x_1^{[m\cdot R_>]}\ =\ (-x_1)^m\ ,\ \ 
x_2^{[n\cdot R_>]}\ =\ (-x_2)^n\ ,\ \ 
x_3^{[l\cdot R_>]}\ =\ (-x_3)^l\ .}
The second of these corresponds to a state of depth $d=n$, and so gives 
$t=d+\nu\cdot\theta= n+(l+m)$. The first and third can be analysed 
using formulas obtained by permuting $\lambda,\mu,\nu$ and 
$x_1,x_2,x_3$ cyclically in \gwnsW. We get $t=d+\mu\cdot\theta = 
m+(n+l)$ from the first, and $t=d+\lambda\cdot\theta = 
l+(m+n)$ from the third monomial of \mnlR. 

From \mnlR, we see that 
\eqn\trixx{
 \matrix{\quad\cr x_1\cr
	x_2~~\quad x_3\cr}
\ }
is the $su(2)$ analogue of the $su(3)$ result \trix.

\listrefs
\bye